\documentstyle[12pt]{article}
\def\lsim{\mathrel{
\lower4pt\hbox{$\sim$}}\hskip-12pt\raise1.6pt\hbox{$<$
}\;}
\def\BAR{\overline}
\def\gsim{\mathrel{\lower4pt\hbox{$\sim$}}
\hskip-10pt\raise1.6pt\hbox{$>$}\;}

\begin{document}
\vspace*{-.5in}
\rightline{BNL-HET 98/47}
\rightline{AMES-HET-98-15}  

\begin{center}

{\large\bf General analysis of $B$ decays to two pseudoscalars for EWP,
rescattering and color suppression effects}

\vspace{.3in}

David Atwood$^{1}$\\
\noindent Department of Physics and Astronomy, Iowa State University, 
Ames,
IA\ \ \hspace*{6pt}50011\\
\medskip
and\\
\medskip
Amarjit Soni$^{2}$\\
\noindent Theory Group, Brookhaven National Laboratory, Upton, NY\ \
11973\\
\footnotetext[1]{email: atwood@iastate.edu}
\footnotetext[2]{email: soni@bnl.gov}
\end{center}
\vspace{.2in}

\begin{quote}
{\bf Abstract:}  
A general analysis for $B$ decays to two pseudoscalars, involving ten
modes, is presented. A simple model for final state interactions and
rescattering effects is proposed. We show how the data can be used to
deduce important information on electroweak penguins (EWP), rescattering
and color suppression effects and on the CKM parameters in a largely model
independent way by using $\chi^2$-minimization. We find that the current
data suggests the presence of color-suppressed tree at levels somewhat
larger than simple theoretical estimates. 
Once the data improves the extraction of 
$\alpha$ and/or $\gamma$ may become feasible
with this method as we illustrate with the existing data. 
\end{quote}
\vspace{.3in}

The recent CLEO results~\cite{newdata,roy} of $B$ mesons decaying into two
pseudoscalars are significant in that they give strong evidence of QCD
penguin processes, in particular in the instance of $K\pi$ final states
where penguins are thought to dominate. 
Of course in the amplitude to $K\pi$ final states, the tree decay
$b\to u\BAR u s$
will also give a contribution which will interfere with the penguin
decays.

Evidence for the influence of the tree graph may be found by considering 
the ratio:

\begin{equation}
R_K = Br(\BAR B^0\to \BAR K^0\pi^0) / Br(\BAR B^-\to \BAR K^0\pi^-)
\end{equation}

This has the value $R_K\approx 0.81\pm 0.35$ where the expectation if
only penguin processes were involved would be $1/2$. Clearly, no firm
conclusion can be drawn from this central value due to the large error;
however, since the color allowed tree does not contribute to either the
numerator or denominator of this ratio, it does seem to suggest that the
color suppressed tree (CST), rescattered tree (RST) 
and/or electroweak penguins (EWP) may
play an important role.

These recent experimental findings are attracting considerable theoretical
attention \cite{neubert, deshpande, gronau, buras}.  The goal of this
Letter is to propose a simple model which includes all of the features of
the standard model and can be fit to current and future data. This will
point to the possible interpretation of the current data and point out
patterns that should be examined in future experiments.  Needless to say
it is very important to acquire a quantitative understanding of the role
of these individual contributions such as CST, EWP and RST as
they enter in an intricate manner in our ability to deduce the CKM
parameters from the experimental data and to test the Standard Model (SM)
through the use of the unitarity triangle for the presence of new
physics---a goal of unquestionable importance.

Bearing all that in mind and also that large increases in the data sample
are expected in the near future we will attempt to construct a general
framework for analysis of these and related issues. We want to focus on
the ten modes of $B$ decays to two pseudoscalars, as these are the
simplest, for which CLEO has already presented some data
\cite{newdata,roy}: 1) $K^-\pi^0$, 2) $K^0\pi^-$, 3) $K^0\pi^0$, 4)
$K^-\pi^+$, 5) $\pi^-\pi^0$, 6) $\pi^+\pi^-$, 7) $\pi^0\pi^0$, 8)
$K^-K^0$, 9) $K^+K^-$, 10) $K^0\BAR K^0$. The amplitudes ($M_i$ for
$i=1\dots10$) for these reactions receive contributions from four Standard
Model processes: Tree (i.e. the color allowed piece), final state
interactions (FSI) or rescattered tree (RST), color suppressed tree (CST),
penguin and EWP.  As is well known, at present no first principles method
exists for calculating final state interactions or rescattering effects.
We propose a simple model for that purpose. We propose that all FS
rescattering effects are due to quark level reactions of the type $u\BAR u
\to u\BAR u$, $d\BAR d$, $s\BAR s$ etc.  The characteristic strength of
such a conversion, presumably of some non-perturbative origin, will be
denoted by a complex parameter $\kappa$ which, for simplicity, is taken to
respect exact SU(3) flavor symmetry.

For each of the ten reactions we first
decompose the amplitudes in terms of
the quark level processes:

\begin{eqnarray}
M_i
=V_{tq}^*V_{tb}P_i
+V_{uq}^*V_{ub}T_i
+V_{uq}^*V_{ub}\hat T_i
+V_{tq}^*V_{tb}E_i
\label{su3_a}
\end{eqnarray}

\noindent
where $P_i$ is the penguin contribution to the amplitude,  $T_i$ is the
tree, $\hat T_i$ is the color suppressed tree, and $E_i$ the 
electro-weak
penguin. In the above, $q=s$ in the case of final states which carry
one unit of net strangeness (i.e.\ $K\pi$) and $q=d$ in the cases with
no net strangeness ($\pi\pi$ and $K\BAR K$).  Note also
that we have used unitarity
to eliminate terms proportional to $V_{cq}^*V_{cb}$. In this notation,
therefore, the difference between graphs with an internal $u$-quark and 
an
internal $c$-quark contributes to what we designate the tree. This 
causes no
trouble since such contributions have
SU(3) properties consistent with
part of the tree amplitude and this notation is consistent with what is
usually taken as tree-like processes~\cite{charles}.

Using the conventions of~\cite{zeppenfeld}, SU(3)
provides the following relationships between these amplitudes:

\begin{eqnarray}
&P_1=-P_2/\sqrt{2}=-P_3=P_4/\sqrt{2};~~~P_8=
-P_1/\sqrt{2}
\nonumber\\
&P_9=\sqrt{2}P_1+P_6;~~~P_5=0;~~~P_6=-\sqrt{2}P_7=-P_{10}
\nonumber\\
&T_3=T_1+T_2/\sqrt{2}-T_4/\sqrt{2};~~~T_5=T_1+T_2/\sqrt{2};
\nonumber\\
&T_7=T_1+T_2/\sqrt{2}-T_6/\sqrt{2};~~~T_8=T_2;~~~T_9=-T_4+T_6;
\end{eqnarray}

\noindent Note that  the relations given above for $T_i$ apply also to
$\hat T_i$ and $E_i$. To specify our model, we therefore need
give the expressions only for
$\{P_1, P_6\}$ and $\{T_1,T_2,T_4,T_6,T_{10}\}$ (and likewise for
$\hat T$ and $E$) whence the other terms are determined.

Let us now introduce the quantities $p$, $t$, $\hat t$, $e$ as the
basic penguin, tree, color  suppressed tree and EWP amplitudes  at the
quark level.  In our rescattering model,  the meson decay amplitudes
can be expressed as:

\begin{eqnarray}
&P_1=(1+3\kappa)p/\sqrt{2};
~~~
P_6=(1+5\kappa)p;
\nonumber\\
&T_1=(1+\kappa)t/\sqrt{2};
~~~
T_2=-\kappa t;
~~~
T_4=(1+\kappa)t;
~~~
T_6=(1+2\kappa)t;
~~~
T_{10}=-\kappa t;
\nonumber\\
&\hat T_1=(1+\kappa)\hat t/\sqrt{2};
~~~
\hat T_2=-\kappa \hat t;
~~~
\hat T_4=\hat T_6=\hat T_{10}=0;
\nonumber\\
&E_1=(1+\kappa/3)e/\sqrt{2};
~~~
2E_2=E_4=E_6=-E_{10}=-2\kappa e/3;
\end{eqnarray}

In this model we take the quark level quantities to be pure real and
attribute all rescattering phases to the quantity $\kappa$. The model,
therefore, for simplicity, does not include, for now,  phases which can
occur at the  quark level
in perturbation theory calculations~\cite{directcp}.

It is easy to see that our calculational procedure, in general, involves
the following eight parameters: $t$, $\hat t$, $p$, $e$, Real
$\kappa\equiv\kappa_R$, Im $\kappa\equiv\kappa_I$, $\rho$ and $\eta$ where
$\rho$ and $\eta$ are the CKM parameters \cite{gamma} in the Wolfenstein
parameterization which are rather poorly determined. Indeed a recent fit
gives \cite{mele}, $\rho=.10^{+.13}_{-.38}$, $\eta=.33^{+.06}_{-.09}$. Of
course, $\lambda\equiv \sin\theta_c \simeq .22$ and $A\simeq .81$ are the
other two CKM parameters that enter these decays. We will take the results
of this fit of the CKM parameters as an input to our subsequent fits.

In the model there are, in general, 20 reactions which are controlled by a
maximum of eight parameters. The 20 reactions are the ten listed above for
$B^-$ and $\BAR B^0$ decay together with their charge conjugate 
decays in the case of $B^+$ and $ B^0$ decay.

In current data, the conjugate pairs have been taken together since in
these modes CP violation has not been experimentally measured. If one
takes the modes averaged with their conjugates, then, for now, we can
regard these as ten reactions controlled by a maximum of eight parameters.
We will search for self consistent solutions to the data and solve for
these parameters along the way by using $\chi^2$-minimization. As an
illustration of how the model works we will use the currently available
experimental data. Clearly as the quality of the data improves one can
hope to improve the evaluation of these parameters as well.  Improvements
will also result when CP violating rate differences are either measured or
bounded in these modes. The $\chi^2$ function we find is flat in the
direction of changing $\eta$ suggesting that without measurements of CP
violation, the existing data is unable to provide a useful constraint on
$\eta$. In our fits, we thus hold $\eta$ fixed at the central value
of~\cite{mele}; thus for analyzing the current data averaged over
conjugate pairs we have in effect only 7 parameters for the ten modes.

The input data we use in our fits is shown in Table~\ref{tabzero}.
We systematically study eight types of fits to this data
(see Table~\ref{tabone}) to search for
the presence of EWP, RST and CST i.e.\ to solve for $e$, $\kappa$ and
$\hat t$ (along with the other parameters). In this table, we use the
notation 
$     t_0=|V_{us}^*V_{ub}| t$,
$     p_0=|V_{ts}^*V_{tb}| p$ and
$     e_0=|V_{ts}^*V_{tb}| e$.

Solution $A$ is with $e$, $\hat t$ and $\kappa$ switched off (i.e.\
$e=\kappa= \hat t=0$) and thus has the worst $\chi^2$
amongst the eight cases.

In particular the best fit under this hypothesis has low $Br(\BAR
B^0\to \BAR K^0\pi^0$) compared to experiment (Table~\ref{tabzero}). As we
shall see, this trend continues in the fits with more free parameters. 
Also the ratio $R^-_\pi \equiv Br(B^-\to \pi^-\pi^0)/Br(B^0
\to\pi^+\pi^-)$ has the experimental result $R^-_\pi\approx 1.15\pm
0.6$ while the fit gives 0.64. Although this  in itself is barely
significant, again it is interesting to follow this ratio through the
other cases.

Set $B$ consists of three solutions ($B1, B2, B3$) which allow switching
on of either $e$, or $\kappa$ or $\hat t$ respectively. Thus, for example,
for $B1$, $\kappa=\hat t = 0$; and $\chi^2$-minimization is used to solve
for $e$. Similarly for $B2$, $e=\hat t=0$ and $\kappa\ne0$ and for $B3$,
$e=\kappa=0$ and $\hat t \ne0$.  Amongst these three solutions in the $B$
set, the set $B3$, where the CST is turned on, is the best fit. Indeed it
has the highest confidence level of all of the solutions.  Likewise the
solution $B2$ which allows rescattering has a high confidence level. This
suggests that the fit of the basic model $A$ is best improved by allowing
a large CST contribution or introducing rescattering.  In the case $B2$,
where $\kappa$ is introduced, we obtain a satisfactory fit which, however,
has the important consequence that it 
gives a substantial phase to $\kappa$ which in turn could give
rise to substantial CP violation.

Set $C$ (consisting of three solutions) requires two of the three types of
contributions to be non-vanishing simultaneously,  
while the set $G$ allows
CST, FSI and EWP to be all present simultaneously.

It is interesting to compare $C3$ where the CST is turned off, in other
words where we assume that the data is explained with rescattering and EWP
with $B3$ where only the CST only is turned on. It would seem that these
two complimentary models are best distinguished by considering the
$\pi\pi$ and $KK$ modes. First of all, in $B3$,
$Br(\pi^-\pi^0)>Br(\pi^+\pi^-)$ while in the case of $C3$ they are roughly
the same. The $\pi^0\pi^0$ mode is smaller in the case of $C3$ and, by
considering all the cases it is apparent that this mode is particularly
sensitive to the presence of the CST.  On the other hand, the $K^-K^0$
mode is larger in the case of $C3$ since, as has been pointed out in the
literature~\cite{lipkin,PRD_58,mannel}, this mode is sensitive to
rescattering effects because the tree cannot contribute to this mode
without rescattering. The same is also true for the mode $K^+K^-$ where
the quark content of the final state can only arise through rescattering.

Of course none of the hypotheses are in any way ruled out or clearly
favored by the current data; however, since the largest confidence level
is for solution $B3$, it does suggest that perhaps color-suppression does
not hold too well for the final states with two light pseudoscalars (i.e.\
$\pi\pi$, $K\pi$ and $K\BAR K$) that are under consideration. Indeed in
all the instances where the CST is allowed, $\hat t/t\approx 1$ whereas
naive color counting give $\hat t/t=1/3$.  Alternatively, as shown by
solution $B2$ rescattering provides an adequate explanation for the data
although it becomes difficult to make $\pi^-\pi^0$ larger than
$\pi^+\pi^-$.

It is interesting to note that in all of the the fits, the trend is to 
consistently give a value for the branching ratio to $\BAR K^0\pi^0$ 
which is about one standard deviation below the current central value. 
It would be interesting if the central value remained at this 
level. For example, if one assumes that the error is reduced to 
$\pm 0.15$ while holding the central value and all the rest of the data 
fixed, Solution $A$ and $B2$ begin to become untenable with confidence 
levels of 0.07 and 0.09. In this scenario $B3$ and $C3$ are the best 
solutions with confidence levels 0.30 and 0.49 respectively
and so CST or EWP with RST would be the favored explanations.

We can also generalize the model somewhat by allowing the rescattering of
the form $q_i\BAR q_i\to q_i \BAR q_i$ to be different from $q_i\BAR
q_i\to q_j \BAR q_j$ where $i\neq j$.  In particular, let us replace
$\kappa$ with $\kappa+\delta\kappa$ in the case where $q_i\BAR q_i\to q_i
\BAR q_i$.  In the solution $g$ listed in the Table we assume that
$\delta\kappa$ is purely imaginary.  Clearly the values of the resulting
fit are similar to the case $G$ where $\delta\kappa=0$. 
And since the confidence level is only 0.33, this does not appear to be a 
particularly useful parameter to explain the current data.

We now briefly comment on some of the  other interesting
implications of these two solutions (see Table~\ref{tabtwo}):

\begin{enumerate}

\item A primary objective of the intense experimental and
theoretical studies of B decays is to deduce the CKM phases
of the unitarity triangle. In this regard, as an illustration,
we note from Table 3 that solutions to the current data
with $CL \gsim 0.6$ seem to give $ \gamma = 90 \pm 20$ degrees.

\item The ratio ``penguin-like''/``tree-like'' \cite{charles} for the
$\pi^+\pi^-$ is a useful parameter for facilitating the extraction of
$\alpha$ from the measurement of the time dependent CP asymmetry in
$B\to\pi^+\pi^-$ following the works of Charles \cite{charles} and of
Quinn and Grossman \cite{gross}.  We see (Table~3) that for the various
solutions considered, this ratio ($p_\pi/t_\pi$) is
$\sim$0.3--0.6.  Note also that in the analysis of
Ref.~\cite{charles,gross} EWP contribution was assumed negligible.  We
find the EWP contribution to be about 1\% of the total amplitude for the
$\pi^+\pi^-$ mode---thus quite small. So their idea for extraction of
$\alpha$ has a good chance of working if an accurate determination of the
penguin/tree for the $\pi\pi$ mode can be achieved.

\item The current fit to the model suggests that the EWP 
effects are small. However, since the possible presence of the EWP in
these pseudoscalars modes 
can adversely affect
a model independent determination of
$\alpha$ and/or $\gamma$, $B$ decays to two vector modes \cite{atwood}
become an important check of this.  These modes (e.g.\ $K^\ast\omega$,
$K^\ast\rho$ and $\rho\omega$) allow a quantitative determination of EWP
in a model independent way and those in which EWP are not prominent (e.g.\
are color suppressed) can then be used for deduction of $\alpha$ or
$\gamma$.

\item In conjunction with the assumption of flavor SU(3) for the
rescattering effects, the method used here allows one to incorporate also
the information from $B_s$ decays, e.g.\ $B_s\to K^+K^-$, $K^0\BAR K^0$,
$K^+\pi^-$, $K^0\pi^0$, etc, in the determination of the parameters of
interest. Hopefully, these measurements would become accessible at
various accelerators in the not too distant future.

\item We have examined partial rate asymmetry (PRA) ( as usual defined to be
$[\Gamma(\BAR B\to X)-\Gamma(B\to \BAR X)] /[\Gamma(\BAR B\to
X)+\Gamma(B\to \BAR X)]$) in these ten modes for the solutions which allow
$Im(\kappa)$ to be non-zero where we are taking the needed strong phase
originating only from $Im(\kappa)$ characteristic of the RST\null.  Since
the likelyhood distribution for these quantities is not well described by
a Gaussian, the number we give indicates the range of values for the
magnitude of the PRA which contains 68\% of the
likelyhood distribution with the remainder split evenly above and below
the range.  The absolute sign of the PRA, of course,
is proportional to the sign of $Im(\kappa)$ which 
cannot be determined without some CP
odd experimental data. 

\indent
We find that the PRA in the $K\pi$ modes tend to
be $\lsim 15\%$ whereas for the $\pi\pi$ modes appreciably larger
asymmetries $\sim10$--80\% seem possible \cite{directcp} if indeed 
large $Im(\kappa)$ is possible.

Measurments of direct CP violation in the $\pi \pi$ final states
will, of course, be useful in clarifying whether there is
large rescattering present in B decays to two
pseudoscalars.  
The same is
also true for $K\pi$ final states, although the CP violation is smaller, 
it should be easier to measure.

\item
Perhaps the most interesting finding suggested by this study is that 
in contrast to $B\to D$ decays, the best fit to the data in our model 
seems to suggest that there is little or no color suppression operative in 
$B$ decays to two pseudoscalars. 
It will be extremely interesting if 
this
assessment persists with improvements in the data.

\item In all the solutions the central value for the EWP is small,
comparable with theoretical estimates~\cite{fleisher,wilson}. Clearly there
is considerable uncertainty and since the impact of EWP is larger in the
$K\pi$ modes, improved data on those modes should clarify the situation. 
For example, if we perform the same fit with the branching ratio for
$\pi^0\BAR K^0$ increased to $2\times 10^{-5}$ (i.e. 1-sigma) and all the
other data and error bars fixed, then fit G produces a value for $e_0$
about 1/2 that of $p_0$ with a confidence level of 0.30. In this case, the
best solution is $C3$ (EWP and rescattering turned on) with a confidence
level of 0.49 with a similar value for $e/p$.

\item Finally, since $\pi\pi$ and $KK$ states receive a larger influence
from the tree graph, one would expect more data on these processes to be
helpful in distinguishing between CST and RST explanations. Indeed the
results in Table~1 suggest how to interpret experimental results to
determine whether CST or rescattering are present.  As discussed above the
presence of the CST is associated with larger branching ratios to
$\pi^0\pi^0$ while the presence of rescattering is associated with larger
branching ratios to $K^-K^0$ and $K^+K^-$.

\end{enumerate}

We are grateful to George Hou for discussions.  This research was
supported in part by US DOE Contract Nos.\ DE-FG01-94ER40817 (ISU) and
DE-AC02-98CH10886 (BNL).

\def\tabstrut{\vrule height13pt  depth2pt width0pt}

\begin{table}[h]
\begin{center}
\caption{
Summary of the experimental data that we use in constructing the fits to
our model.  The branching ratios are in units of $10^{-5}$ and are the
average for each given mode and the conjugate.
The results in the case of
$K^-\pi^+$,
$K^-\pi^0$
and
$\BAR K^0\pi^+$
are the experimental results from~\cite{newdata,roy}.
In the other cases, we infer the branching ratios with the stated errors
from the yields together with the efficiencies given in~\cite{roy}.
\label{tabzero}
}
\medskip
\end{center}
\begin{center}
\begin{tabular}{|c|c|c|}
\hline
\tabstrut Mode \# & Mode & $Br$ \\
\hline
\tabstrut~1 & $K^-\pi^0$ &                 $1.21^{+.30}_{-.28}$ \\
\tabstrut~2 & $\BAR K^0\pi^-$ &            $1.82^{+.46}_{-.40}$ \\
\tabstrut~3 & $\BAR K^0\pi^0$ &            $1.48^{+.59}_{-.51}$ \\
\tabstrut~4 & $K^-\pi^+$ &                 $1.88^{+.28}_{-.26}$ \\
\hline
\tabstrut~5 & $\pi^-\pi^0$ &               $0.54^{+.25}_{-.26}$ \\
\tabstrut~6 & $\pi^+\pi^-$ &               $0.47^{+.15}_{-.18}$ \\
\tabstrut~7 & $\pi^0\pi^0$ &               $0.16^{+.16}_{-.10}$ \\
\hline
\tabstrut~8 & $K^-K^0$ &                   $0.20^{+.24}_{-.16}$ \\
\tabstrut~9 & $K^+K^-$ &                   $0\pm   .5$ \\
\tabstrut10 & $K^0\BAR K^0$ &              $0\pm  1.7$ \\
\hline
\end{tabular}
\end{center}
\end{table}

\begin{table}
\begin{center}
\caption{Summary of the solutions under the various hypothesis. $N$
means the corresponding entity is switched off (i.e.\ set to zero)
whereas $Y$ means it is on and being determined. The branching ratio
for the ten modes are given in units of $10^{-5}$; the individual
amplitudes $t_0$, $p_0$, etc.\ are expressed accordingly.\label{tabone}}
\medskip
\begin{tabular}{|c|r|r|r|r|r|r|r|r|r|}
  \hline
  ~ & $g$ & $G$  & $C3$ & $C1$ & $C2$ & $B3$ & $B1$ & $B2$ & $A$  \\
  \hline 
  \hline
$\hat t$        & Y& Y&    
N&    
Y&    
Y&    
Y&    
N&    
N&    
N\\   
\hline 
$e$             & Y & Y&    
Y&    
N&    
Y&    
N&    
Y&    
N&    
N\\   
\hline 
$\kappa$        & Y& Y&    
Y&    
Y&    
N&    
N&    
N&    
Y&    
N\\   
\hline 
$\delta\kappa$    & Y& N&    
N&    
N&    
N&    
N&    
N&    
N&    
N\\   
\hline 
  \hline 
$t_0$             &    -0.01 & 0.11&    
   -0.28&    
   -0.11&    
    0.14&    
   -0.15&    
    0.23&    
   -0.28&    
    0.23\\   
\hline 
$p_0$             &    -0.19 & 0.80&    
   -2.21&    
    0.87&    
   -1.42&    
    1.41&    
    1.37&    
   -2.21&    
    1.37\\   
\hline 
$\hat t_0$        &     0.07 &0.15&    
    0.00&    
   -0.16&    
    0.13&    
   -0.13&    
    0.00&    
    0.00&    
    0.00\\   
\hline 
$e_0$             &     0.04 & 0.06&    
   -0.02&    
    0.00&    
   -0.04&    
    0.00&    
   -0.03&    
    0.00&    
    0.00\\   
\hline 
$Re(\kappa)$    &     -0.15 & 0.27&    
   -0.16&    
    0.20&    
    0.00&    
    0.00&    
    0.00&    
   -0.16&    
    0.00\\   
\hline 
$Im(\kappa)$    &     -1.08 & 0.00&    
   -0.12&    
    0.00&    
    0.00&    
    0.00&    
    0.00&    
   -0.12&    
    0.00\\   
\hline 
$\rho$          &     0.13 & 0.08&    
   -0.02&    
    0.04&    
   -0.04&    
   -0.01&    
   -0.08&    
   -0.02&    
   -0.07\\   
\hline 
$\eta $         &     0.32 & 0.33&    
    0.33&    
    0.33&    
    0.33&    
    0.33&    
    0.33&    
    0.33&    
    0.33\\   
\hline 
  \hline 
$K^-\pi^0$      &     1.05 & 1.06&    
    1.02&    
    1.06&    
    1.06&    
    1.02&    
    0.99&    
    1.01&    
    1.03\\   
\hline 
$K^0\pi^+$      &     1.97 & 2.03&    
    1.92&    
    1.95&    
    2.01&    
    2.00&    
    1.87&    
    1.92&    
    1.88\\   
\hline 
$K^0\pi^0$      &     1.18 & 0.98&    
    0.94&    
    0.95&    
    0.98&    
    1.01&    
    0.98&    
    0.96&    
    0.94\\   
\hline 
$K^-\pi^+$      &     1.96 & 1.95&    
    2.02&    
    1.99&    
    1.98&    
    2.00&    
    2.08&    
    2.02&    
    2.06\\   
\hline 
$\pi^-\pi^0$    &     0.57 & 0.57&    
    0.50&    
    0.57&    
    0.60&    
    0.60&    
    0.34&    
    0.50&    
    0.34\\   
\hline 
$\pi^-\pi^+$    &     0.47 & 0.46&    
    0.50&    
    0.46&    
    0.45&    
    0.45&    
    0.52&    
    0.50&    
    0.53\\   
\hline 
$\pi^0\pi^0$    &     0.16 & 0.16&    
    0.10&    
    0.15&    
    0.14&    
    0.14&    
    0.04&    
    0.10&    
    0.04\\   
\hline 
$K^-K^0$        &     0.20 & 0.20&    
    0.19&    
    0.20&    
    0.12&    
    0.11&    
    0.12&    
    0.19&    
    0.12\\   
\hline 
$K^+K^-$        &     0.01 & 0.02&    
    0.07&    
    0.02&    
    0.00&    
    0.00&    
    0.00&    
    0.07&    
    0.00\\   
\hline 
$K^0K^0$        &     0.29 & 0.17&    
    0.15&    
    0.18&    
    0.12&    
    0.11&    
    0.12&    
    0.15&    
    0.12\\   
\hline 
  \hline 
$\chi^2$        &     0.96 & 1.62&    
    2.77&    
    1.65&    
    2.10&    
    2.13&    
    4.68&    
    2.78&    
    4.71\\   
\hline 
$df$            &   1 & 2&    
  3&    
  3&    
  3&    
  4&    
  4&    
  4&    
  5\\   
\hline 
$\chi^2/df$     &     0.96 & 0.81&    
    0.92&    
    0.55&    
    0.70&    
    0.53&    
    1.17&    
    0.69&    
    0.94\\   
\hline 
CL              &     0.33 & 0.45&    
    0.43&    
    0.65&    
    0.55&    
    0.71&    
    0.32&    
    0.60&    
    0.45\\   
\hline 
\end{tabular}
\end{center}
\end{table}

\begin{table}
\begin{center}
\caption{Properties of the solutions that appear most relevant to
the CLEO data. Numbers after $K^-\pi^0$,
$\BAR K^0\pi^+\dots\pi^0\pi^0$ are the magnitudes of the partial rate
asymmetry for $B^-$ 
or 
$\BAR B^0$ to the given mode\cite{directcp}. For each mode, the range 
quoted 
encompasses 68\% of the likelihood function with the rest split above and 
below the range.
(See also caption to Table~\protect\ref{tabone}). \label{tabtwo}}
\medskip

\begin{tabular}{|c|r|r|r|r|r|}
  \hline
  ~ & $G$  & $C_3$ & $C_1$ & $B_3$ & $B_2$  \\
  \hline 
  \hline 
$\hat t$        & Y&    
N&    
Y&    
Y&    
N\\   
\hline 
$e$             & Y&    
Y&    
N&    
N&    
N\\   
\hline 
$\kappa$        & Y&    
Y&    
Y&    
N&    
Y    \\   
\hline 
$\kappa_{D}$    & N&    
N&    
N&    
N&    
N\\ 
\hline
\hline
$t_0$             &     0.11$\pm$    0.03&    
   -0.28$\pm$    0.18&    
   -0.11$\pm$    0.03&    
   -0.15$\pm$    0.03&    
   -0.28$\pm$    0.11    
\\   
\hline 
$p_0$             &     0.80$\pm$    0.26&    
   -2.21$\pm$    5.47&    
    0.87$\pm$    0.38&    
    1.41$\pm$    0.07&    
   -2.21$\pm$    3.17    
\\   
\hline 
$\hat t_0$        &     0.15$\pm$    0.03&    
    0.00$\pm$    0.00&    
   -0.16$\pm$    0.03&    
   -0.13$\pm$    0.03&    
    0.00$\pm$    0.00    
\\   
\hline 
$e_0$             &     0.06$\pm$    0.14&    
   -0.02$\pm$    0.19&    
    0.00$\pm$    0.00&    
    0.00$\pm$    0.00&    
    0.00$\pm$    0.00    
\\   
\hline 
$Re(\kappa)$    &     0.27$\pm$    0.18&    
   -0.16$\pm$    0.23&    
    0.20$\pm$    0.21&    
    0.00$\pm$    0.00&    
   -0.16$\pm$    0.14    
\\   
\hline 
$Im(\kappa)$    &     0.00$\pm$    1.00&    
    0.12$\pm$    0.59&    
    0.00$\pm$    1.12&    
    0.00$\pm$    0.00&    
   -0.12$\pm$    0.37    
\\   
\hline 
$\rho$          &     0.08$\pm$    0.16&    
   -0.02$\pm$    0.26&    
    0.04$\pm$    0.15&    
   -0.01$\pm$    0.12&    
   -0.02$\pm$    0.18    
\\   
\hline 
  \hline 
$\gamma$        &    
   76$\pm$   25&    
   93$\pm$   45&    
   83$\pm$   25&    
   92$\pm$   20&    
   94$\pm$   31    
\\   
\hline 
$p_\pi/t_\pi$   &     0.51$\pm$    0.09&    
    0.37$\pm$    0.88&    
    0.59$\pm$    0.23&    
    0.51$\pm$    0.18&    
    0.37$\pm$    0.60    
\\   
\hline 
$e_\pi/A_\pi$   &     0.01$\pm$    0.02&    
    0.00$\pm$    0.04&    
    0.00$\pm$    0.00&    
    0.00$\pm$    0.00&    
    0.00$\pm$    0.00    
\\   
\hline 
  \hline 
$K^-\pi^0$      &       
  .022---.157&    
  .031---.176&    
  .021---.156&
0&    
  .027---.169
\\   
\hline 
$K^0\pi^+$      &       
 .012---.081&    
 .016---.090&    
 .011---.079&    
0&    
 .014---.088 
\\   
\hline 
$K^0\pi^0$      &      
 .019---.133&    
 .016---.091&    
 .018---.129&    
0&    
 .014---.088
\\   
\hline 
$K^-\pi^+$      &     
  .015---.125&    
  .031---.173&    
  .015---.127&    
0&    
  .028---.170
\\   
\hline 
$\pi^-\pi^0$    &       
  0&    
  0&    
  0&    
0&    
  0   
\\   
\hline 
$\pi^-\pi^+$    &     
 .073---.594&    
 .127---.749&    
 .070---.591&    
0&    
 .110---.719

\\   
\hline 
$\pi^0\pi^0$    &      
  .198---.849&    
  .324---.900&    
  .192---.854&    
0&    
  .291---.902
\\   
\hline 
\end{tabular}

\end{center}

\end{table}


\newpage




\begin{thebibliography}{99}

\bibitem{newdata} R.~Poling, XIX International Symposium on
Lepton and Photon Interactions at High Energies, 
Stanford University (Aug. 1999).


\bibitem{roy} See the talks by J. Roy and J. Alexander at the ICHEP98,
Vancouver (July 1998). 



\bibitem{neubert}
M. Neubert and J.L. Rosner, Phys. Rev. Lett. {\bf 81}, 5079 (1988);
M.~Neubert and J.~L.~Rosner, Phys. Lett {\bf B411}, 403 (1998);
M.~Neubert, JHEP {\bf 02}, 014 (1999).


\bibitem{deshpande} N. G. Deshpande, X.-G. He, W.-S. Hou
and S. Pakvasa, Phys. Rev. Lett. 82, 2240 (1999). 


\bibitem{gronau} M. Gronau and J.L. Rosner, Phys. Rev. D59, 113002, 1999.  

\bibitem{buras} A.J. Buras and R. Fleischer, hep-ph/9810260. 


\bibitem{charles}
For the precise definitions of the phrases ``penguin-like'' and 
``tree-like''
see:
J. Charles,
Phys. Rev. {\bf D59}, 054007 (1999).


\bibitem{zeppenfeld} D.~Zeppenfeld, Z. Phys. {\bf c8}, 77 (1981). See
also reference~\cite{rosneretc}.


\bibitem{directcp} 
Note that the PRAs for these solutions
exhibit the correlations found in similar studies; see, in particular,
M. Neubert, JHEP {\bf 02}, 014 (1999) and D. Atwood and A. Soni, Phys.\
Rev.\ {\bf D58}, 036005 (1998).
Note also that in the estimates for the PRAs given
in Table~\ref{tabtwo} we have, for 
simplicity ignored
the ``perturbative phase'' (see M. Bander, D. Silverman and A. Soni,
Phys.\ Rev.\ Lett.\ {\bf43}, 242 (1979)) and assumed that the penguin
amplitude is purely real.  


\bibitem{gamma} Recall that $\gamma\equiv
\arg(-V_{ud}V^\ast_{ub}/V_{cd}V^\ast_{cb})$. See, e.g., the review
article by H. Quinn in Review of Particle Properties, R.M. Barnett {\it
et al}., Phys.\ Rev.\ {\bf D54}, 1 (1996). 


\bibitem{mele} S. Mele, hep-ph/9808411.  


\bibitem{lipkin}
H.~Lipkin, Phys. Lett. {\bf B445}, 403 (1999).



\bibitem{PRD_58}
M.~Gronau and J.~L.~Rosner, {\bf D58}, 113005 (1998).



\bibitem{mannel}
R.~Fleischer and T.~Mannel, Phys. Rev. {\bf D57}, 2752 (1998);
R.~Fleischer and T.~Mannel, Nucl. Phys. {\bf B533}, 3 (1998).



\bibitem{gross} Y. Grossman and H.R. Quinn, Phys.\ Rev.\ {\bf D56},
7259 (1997).



\bibitem{atwood} D. Atwood and A. Soni,   
Phys. Rev. {\bf D59}, 013007 (1999). 




\bibitem{fleisher} R. Fleischer, Z. Phys.\ {\bf C62}, 81 (1994); N.G.
Deshpande and X.-G. He, Phys.\ Rev.\ Lett.\ {\bf74}, 26 (1995); M.
Groanu, O.F. Hernandez, D. London and J.L. Rosner, Phys.\ Rev.\ {\bf
D52}, 6374 (1995), A.J. Buras and R. Fleischer, Phys.\ Lett.\ {\bf
B365}, 350 (1996); N.G. Deshpande, X.-G. He and S. Oh, Zeit.\ Phys.\
{\bf C74}, 359 (1997); and D. Atwood and A. Soni, Phys.\ Rev.\ Lett.\
{\bf16}, 3324 (1998). 




\bibitem{wilson} Using the Wilson coefficients (given in G. Buchalla,
A.J. Buras and M.E. Lautenbacher, Rev.\ Med.\ Phys.\ {\bf 68}, 1125
(1996)) corresponding to $\Lambda^{(5)}_{\rm QCD}= 140, 225, 310$ MeV,
in the NDR scheme, we find the ratio $e/p=.21,.19$ and .18
respectively.




\bibitem{rosneretc} M.~Gronau, O.~Hernandez, D.~London and J.~Rosner
Phys. Rev. {\bf D52}, 6374 (1995); Phys. Rev. {\bf D52} 6356 (1995).
















\end{thebibliography}
\end{document}